\documentclass[journal]{IEEEtran}

\usepackage[left=0.75in,right=0.75in,top=0.75in,bottom=0.75in]{geometry}

\usepackage{color}
\usepackage{qtree}
\usepackage{times}
\usepackage{epsfig}
\usepackage{graphicx}
\usepackage{epstopdf}
\usepackage{algpseudocode}
\usepackage{algorithm}
\usepackage{amsmath}
\usepackage{amssymb}
\usepackage{amsxtra}
\usepackage{multirow}
\usepackage{mathtools}
\usepackage{cleveref}
\usepackage{url}
\usepackage{cite}
\usepackage{amsthm}
\usepackage{adjustbox}
\usepackage[latin1]{inputenc}
\usepackage{caption}
\usepackage[caption = false]{subfig}
\usepackage{float}

\begin{document}
%
\title{RIoTS: Risk Analysis of IoT Supply Chain Threats}
%
%
%

\author{ \IEEEauthorblockN{\large Timothy Kieras}, \IEEEauthorblockN{\large Muhammad Junaid Farooq}, and \IEEEauthorblockN{\large Quanyan Zhu} \\ \IEEEauthorblockA{Department of Electrical \& Computer Engineering, Tandon School of Engineering, \\New York University, Brooklyn, NY 11201, USA,} Emails: \{tk2375, mjf514, qz494\}@nyu.edu. \vspace{-0.2in}

}

\maketitle

\begin{abstract}
Securing the supply chain of information and communications technology (ICT) has recently emerged as a critical concern for national security and integrity. With the proliferation of Internet of Things (IoT) devices and their increasing role in controlling real world infrastructure, there is a need to analyze risks in networked systems beyond established security analyses. Existing methods in literature typically leverage attack and fault trees to analyze malicious activity and its impact. In this paper, we develop RIoTS, a security risk assessment framework borrowing from system reliability theory to incorporate the supply chain. We also analyze the impact of grouping within suppliers that may pose hidden risks to the systems from malicious supply chain actors. The results show that the proposed analysis is able to reveal hidden threats posed to the IoT ecosystem from potential supplier collusion.
\end{abstract}


%
\IEEEpeerreviewmaketitle

\vspace{-0.2in}
\section{Introduction}
Ensuring the integrity of the supply chain for ICT equipment is becoming an important concern as components are often manufactured, owned, and operated by different entities across the globe~\cite{supply_chain_integrity}. With the emergence and rapid adoption of IoT technologies in infrastructure systems, these concerns are surging as cyber-physical attacks are becoming more complex and may involve international entities. Supply chain threats pose a striking set of new challenges for the security of IoT systems~\cite{mapping_security_challenge}. In broad terms, supply chain security requires a shift of focus from vulnerability management to robust system modeling. This modeling is the foundation of RIoTS. Existing cybersecurity threat assessment tools such as attack trees are useful starting points to model the threats against IoT systems; however, these tools must be adapted to consider the wider class of potential threats that arise from suppliers. Attack trees are typically constructed from empirical observations of how attackers exploit existing vulnerabilities in a system. By contrast, a supply chain attack is not limited to the exploitation of existing vulnerabilities. Rather, because a supplier might modify any component with very few restrictions, the attack surface of a supply chain threat is coextensive with the entire system. Supply chain threats are therefore a particularly robust type of `unknown unknown'. This qualitatively different type of attack requires risk analysis that uses deep knowledge of the system at each layer of complexity. 


\begin{figure}[t]
    \centering
    \includegraphics[width=.3\textwidth]{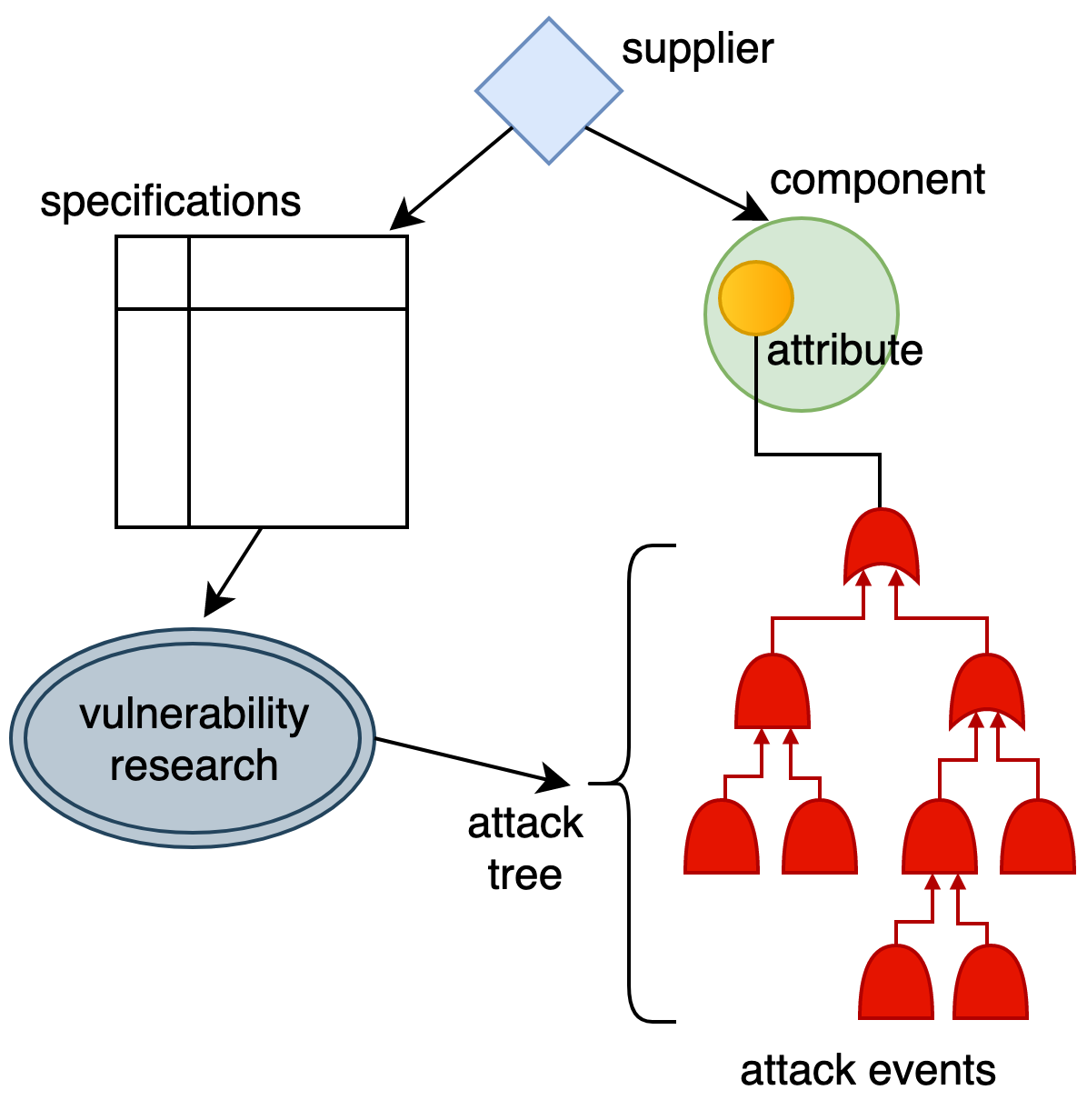}
    \caption{Supplier role in threat assessment. By providing specifications along with components, supplier trust impacts accuracy of attack trees.}
    \label{fig:assessment-process-diagram}
\end{figure} 

\vspace{-0.1in}
\subsection{Related Works}
Security risk modeling techniques and those that use directed acyclic graphs provide necessary background \cite{kordy2014dag, xiong2019threat}. The attack tree method, developed by Schneier and Amoroso, extends from earlier fault tree analysis \cite{RN06973090519990101, amoroso_fundamentals_1994}. Significant developments have been made in the construction and use of attack graphs, including the incorporation of defensive measures, aggregation and probabilistic modeling \cite{roy2012attack, homer2013aggregating, wang2008attack, gribaudo2015exploiting, poolsappasit2011dynamic}. The use of automated attack tree generation at scale is similarly a topic of recent study\cite{xiong2019threat, sheyner2002automated, ou2006scalable}. A more formal analysis of attack trees has also been developed in \cite{mauw2005foundations,jha2002two}. On the other hand, reliability analysis has been investigated in detail in the literature,\cite{rausand2003system, contini2011analysis} with Baiardi et al. studying hierarchical, hypergraph based modeling of systemic security risks \cite{baiardi2009hierarchical}.  Standard definitions of supply chain terminology and discussion of associated risk management practices for ICT systems are found in the report from the National Institute of Standards and Technology (NIST)~\cite{NIST}.

\vspace{-0.1in}
\subsection{Contributions}

A necessary precursor to risk management is the ability to analyze and quantify risk. Traditional methods of risk analysis do not take into account the supply chain layer of the problem. This paper is aimed at developing the foundations for analyzing supplier oriented risks by bridging the gap between established reliability analysis and attack tree security risk assessment techniques. By modeling system hierarchy it is possible to discover supplier involvement and assess hidden risks. We find that this holistic, system level analysis of security risk is essential to approaching supply chain risk. 

The rest of the paper is organized as follows: Section II provides an overview of the essential components used for risk analysis, Section III provides the risk analysis metrics, Section IV presents a case study to evaluate the proposed analysis, and Section V concludes the paper with providing some insights into future work.

\vspace{-0.1in}
\section{Background}
We consider a network of electronic components as an archetype of IoT devices. In this section, we provide an overview of the ingredients that are essential for risk analysis in the IoT system. Supply chain risk modeling comprises three areas of concern, each of which can be modeled as a network: attack steps, supply chains, and component relationships. Understanding supply chain risk involves not only using established techniques of vulnerability assessment, but also a thorough understanding of the social, political and economic context of business interrelationships that are the external reflection of the internal technical structure of the system.

\vspace{-0.1in}
\subsection{Attack Tree}
Attack trees are a graph-based approach to model a set of attack steps with varying logical relations. Originally developed as a variant of fault trees, attack trees are distinguished from fault trees by focusing on security related faults caused by the malicious and intentional activity of some attack agent. A second significant feature of attack trees is that the fault in question may be the disruption or negation of any number of security attributes of the system, a different set of concerns than functionality. Conventionally these attributes are {confidentiality, availability, and integrity}. 

Attack trees are organized around a goal, which is the root node of the tree. Child nodes are subgoals that stand in some logical relationship to their parent. Some nodes are \emph{and} nodes, in which case they require all of their children as preconditions. Other nodes are \emph{or} nodes, such that each of the children represent differing ways to satisfy the node's preconditions. Nodes in an attack tree typically are assigned probability values, making the attack tree a Bayesian network. Given a well developed attack tree and associated probabilities, the likelihood that the attack goal will be reached can be computed. Attack trees often are annotated with other values, such as the impact or cost of an attack step succeeding, the skill level required to accomplish the attack step, or time constraints. While many attack trees are manually constructed by experts in order to conduct a formal risk analysis, in other cases the construction of attack trees can be automated.

\vspace{-0.1in}
\subsection{Supplier Network}
Supply chain risk analysis must ask questions about the relevance of manufacturing across diverse jurisdictions and regulatory regimes. Likewise, concerns about supply chain are recursive, since the chief risk arising from a supplier might be from its own suppliers. In general, these concerns can be abstracted into a trust value, discussed in greater depth in the following subsection. Additionally, difficult questions must be posed about the effect on risk of suppliers being associated by various forms of business relationship, including ownership, investment, partnership, merger or acquisition. In general, these questions will be considered by grouping together certain suppliers under their `owners', although this term is not meant to limit the relationships under consideration to full ownership. Perhaps the most challenging aspect of supply chain security involves the dynamic nature of the supplier network, since a change in the topology of the supplier network may impact the resulting risk in significant ways.

The network of suppliers involved in a system may extend beyond those that immediately produce a component. For instance, a company may own several suppliers as subsidiaries, requiring that the owner be considered as a potential risk to systems that include components manufactured by these subsidiaries. Additionally, as described in the National Institute of Standards and Technology (NIST) framework for supply chain risk management, risk may arise from all entities involved in the entire life-cycle of the system: from design and manufacturing, to transportation, retail, maintenance and even disposal. All of these entities will become part of the risk model.

\vspace{-0.1in}
\subsection{System Graph}
The network of suppliers involved in a system reflects the composition of the system from subsystems and components. Although it is common to consider subsystems as `black boxes' whose internal composition is less relevant than the functions and interfaces exposed by the subsystem, consideration of supplier-induced risk requires examining each layer of complexity. Apparently innocuous components from untrustworthy suppliers may pose significant risks. Generally, a component can be considered either as a component or as a system in its own right. By recursively decomposing a system's components, a more complex network is generated that more fully represents the composition of the system. This sort of decomposition is necessary in order to discover the suppliers involved in a system. As a generalization, each component in a system has a supplier that provides the component to the company that produces the system. As a product becomes incorporated into a larger system, its systemic risk is affected by the suppliers involved in the product's life-cycle.

\begin{figure}[h]
    \centering
    \vspace{-0.1in}
    \includegraphics[width=.3\textwidth]{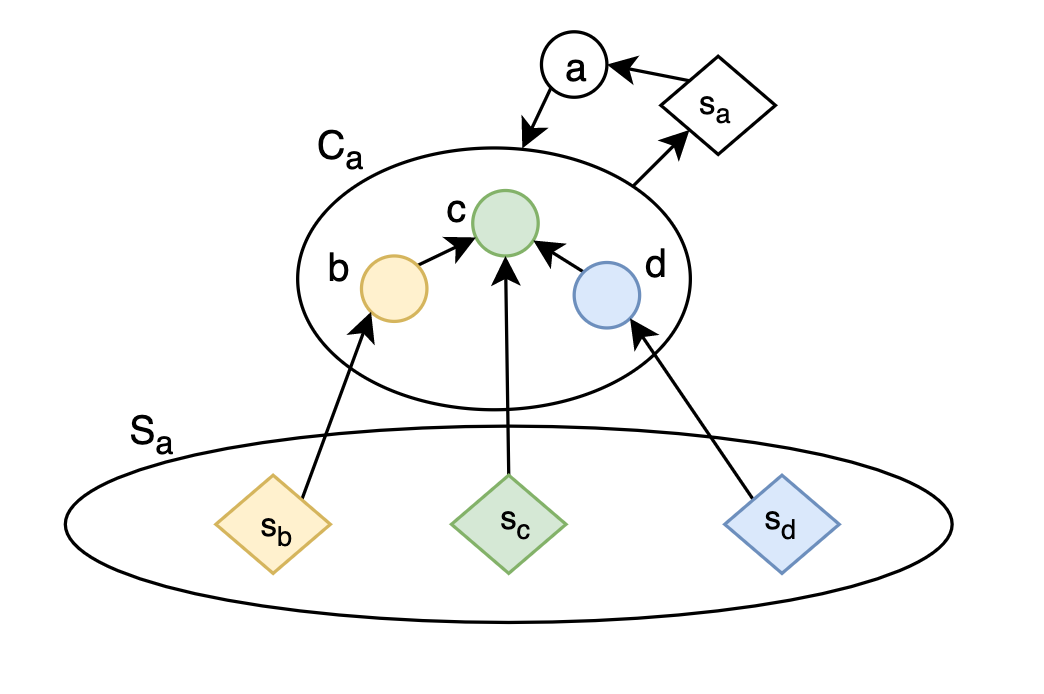}
    \caption{System graph depicting component nodes $C_a$ provided by suppliers $S_a$.}
    \label{fig:sys_graph_example}
\end{figure}

Composing the supplier network together with a graph of component dependencies produces a unified view of the system from the perspective of supply chain security. The resulting graph will be called a system graph. Consulting relevant attack trees for each component yields probability values that will be essential for systemic risk evaluation.  In this way, the risk information contained in an attack tree can be transferred into a graph of the system where components are central rather than attack steps. This shift in focus is essential in order to consider the relationship of suppliers to attack steps.

The following definitions are given to specify the nature of a system graph, with an example shown in Fig.~\ref{fig:sys_graph_example}:
\begin{itemize}
\item \textbf{System Graph}, represented by $SG=(C, S, E, n, u)$, where \(C\) and \(S\) are sets of components and suppliers, respectively, and the set of edges is \(E \subseteq (C \cup S) \times (C \cup S)\).
A component implements a particular set of functions, which is denoted as \(c_f\). Note that a subset of these functions will be relevant to security.
A special node is denoted as \(n \notin C\) that identifies the entire system as a whole. This node is useful when traversing various levels of complexity. 
\item \textbf{Decomposition function}, represented by \(d\) maps a component that is a system to the components out of which the system is composed. Every component in a system can be decomposed into its proper components unless the component is atomic. The function $d$ is defined as: \[\forall c_i \in C_m, \begin{cases} d(c_i) = C_i, \implies c_i = n \in SG_i \\ d(c_i) = \{\emptyset\}\end{cases}\] 
\item \textbf{Component} that is also a system is the product of a composition process conducted by some entity, denoted as \(u \notin S\). The entity conducting the composition process of one system is also the entity that is the supplier when the system is considered as a component. The relation between layers of complexity in a system is defined as follows: \[ \forall c_i \in C_m, c_i, s_i = n, u \in SG_i, \implies \langle s_i, c_i\rangle \in SG_m \]
\item \textbf{Edges} \(e \in C \times C\) indicate functional dependencies, where an edge \( \langle b,c\rangle \) signifies that \(c\) cannot perform its function unless \(b\) is adequately performing its function. In other words, the inputs of function set \(c_f\) are the outputs of function set \(b_f\). Where these functions are relevant to security, these edges encode various ways to attack each component.
\end{itemize}

\vspace{-0.1in}
\subsection{Attack Probabilities and Supplier Trust}
In a system graph as defined above, the risk of any component node being attacked successfully can be assessed with a well-developed attack tree rooted at the component. The intuition to provide similar probability values for supplier nodes requires development. The value should indicate the perceived or assessed trustworthiness of the entity; i.e., the probability that the supplier will not be malicious or compromised. 

Trust and risk are both values that can be interpreted as probabilities. The relation between these values is found in the tacit role that a supplier plays in providing the product specifications that are the foundation of vulnerability assessment. While in some cases empirical techniques exist to discover the functionality of a product, in most cases the supplier is the primary source of this knowledge. In other words, the trust value of a supplier should indicate the degree of belief accorded to the specifications, and therefore to the resulting attack trees. The supplier does not become a node in the attack tree; rather, it generally affects the presumed accuracy of the probabilities in the attack tree.

This definition of supplier trust can be formalized as follows. Let \(F\) be the set of possible functions, and let \(S\) and \(A\) be subsets of \(F\), where \(S\) is the set of stated specifications provided by the supplier, and \(A\) is the set of specifications that the component actually implements. 
\begin{figure}[H]
    \centering
    \includegraphics[width=.3\textwidth]{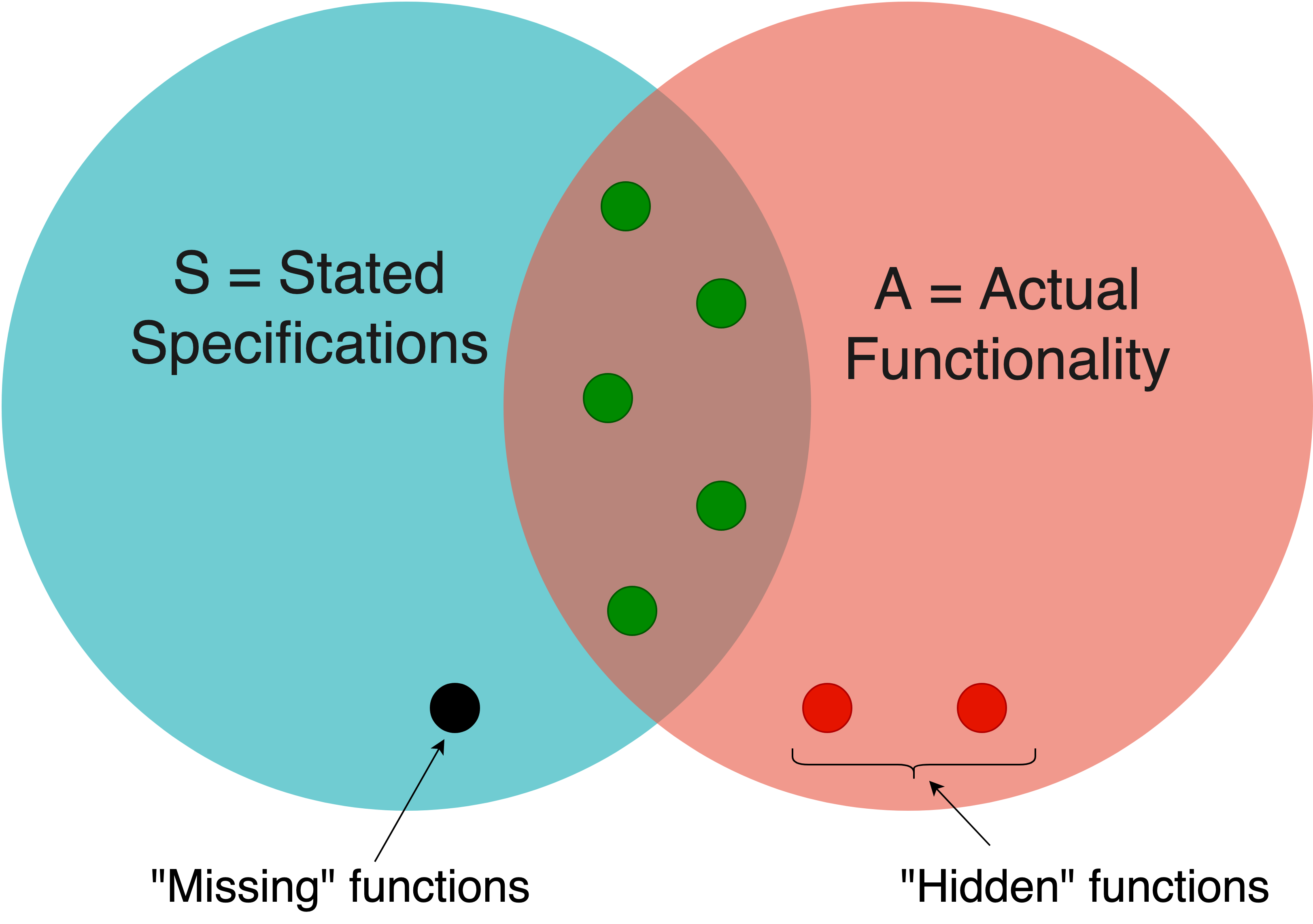}
    \caption{Supplier trust as defined by specification accuracy regarding missing and hidden functions.}
    \label{fig:trust}
\end{figure}
Fig.~\ref{fig:trust} illustrates the two possible ways in which the trust value might be impacted. On the one hand, the actual functionality of the system might include functions not indicated in the specifications. These can be called `hidden' functions. Yet it is not sufficient simply to capture trust as the ratio \(\frac{|A|}{|S|}\), which would be the probability that some actual function \(f_a \) is in \(S\).  It is possible that the specifications may claim a certain function that is not, in fact, implemented -- a `missing' function. To capture this case, we would compute the ratio \(\frac{|S|}{|A|}\), or the probability that some specified function \(f_s\) is in \(A\). To capture both scenarios in one measure, the sample space is defined as \(S \cup A\), or the subset of \(F\) that pertains to the system. An accurately reported specification is one for which \(f_n \in S \implies f_n \in A \text{ and } f_n \in A \implies f_n \in S\), and is indicated as the set \(S \cap A\). Therefore the trustworthiness of a supplier can be measured as the ratio of accurately reported functional specifications to functions that pertain to the system, or
\begin{align*}
\centering
\text{Trust} = \frac{|S \cap A|}{|S \cup A |} \quad\text{where } S, A \subset F\ 
\end{align*}
Interpreting the above ratio as a probability value, supplier trust means the probability that a function pertaining to the system is an accurately reported function. Supplier trust can then be taken into account by considering it as an upper-bound on the accuracy of the attack graphs constructed based on the supplier-provided specifications. If these graphs result in an assessed probability \textit{p} that the system's security will be maintained while under threat, but the specifications are only trusted with trust \textit{t}, then the system security should be assessed as the product of \textit{p} and \textit{t}. The overall relation between supplier, specifications, and threat assessment is shown in Fig.~\ref{fig:assessment-process-diagram}.

\vspace{-0.1in}
\section{Risk Analysis}

In this section we discuss required techniques to analyze risk in a system graph. We focus on the use of minimal cutsets to compute system risk and several measures of component importance. 
\subsection{Minimal Cutset Analysis}
Given the above definitions, the security risk in a system may be approached generically as a problem analogous to system reliability. Before proceeding, it should be noted that this procedure assumes that the system graph has been produced that represents a particular attack scenario. Combining many attack scenarios into a unified system graph is a feasible but complex application of the analysis presented here. While assessing probability of attack success in an attack tree might be conducted by considering the attack tree as a Bayesian network and calculating prior and posterior probabilities of component node failures, this procedure cannot be applied directly to an arbitrary system graph such as described above. This is due to the fact that in a system graph, there may be suppliers with multiple edges to various components such that the risk analysis procedure must take into account this kind of `common cause' failure. Accordingly, the risk analysis conducted here will follow the method of distilling a system graph into a set of minimal cutsets~\cite{rausand2003system}. 

Each cutset consists of a set of components that, if all are successfully attacked, the attack as a whole will succeed. This is analogous to system failure. The cutsets should be minimal, i.e., if one cutset is a subset of another cutset, the latter is not minimal and can be excluded because it represents an attack that includes extraneous steps.


A set of minimal cutsets $A$ is defined so that each cut \(a \in A\) is itself a set of steps defined as \(\lbrace y_i |y_i \in C \cup S\rbrace \), and each step $y$ has a probability of occurring $r_y$. Note that a step $y$ may involve either a supplier or component, at any level of the system. The risk to the system in general can then be calculated using the probability that all of the steps have been achieved of at least one of the cutsets. It can be formally expressed as follows:
\begin{equation}
R(r) = 1-\prod_{a\in A}\left ( 1-\prod_{y\in a}r_y  \right)
\label{equation:risk}
\end{equation}

\subsection{Component Importance Measures}
In reliability theory there are several measures of component importance that are useful for risk analysis in a large system graph. The two that will be discussed here are the Birnbaum Importance (BI) measure and the Improvement Potential (IP) measure. Both measures provide insight into which nodes are significant sources of risk. These insights might be used to investigate possible causes of security incidents, calculate cost-effective mitigation strategies, or assess the security risk of a system during design.

\begin{itemize}
    \item \textbf{Improvement Potential:} In the context of reliability, IP of a node represents the potential gain in system reliability that could be achieved by maximal improvement of the node's reliability. In the context of risk analysis, the goal is risk minimization, and so the improvement potential should be calculated accordingly. Equation \eqref{equation:risk} indicates the required structure function needed to calculate the Improvement Potential measure. Because the effect on risk of any given node depends not only on its own risk value but also the risk values of other nodes in the system, Improvement Potential of a node must be calculated based on some particular vector of risk probabilities. Let two vectors be defined to represent the initial state of the system and the system state after improving the node in question, as follows:
\begin{align*}
& s_i^0= \{r_j | r_j \in [0,1]  \text{ for } j \in 1..N \} \\
& s_i^1= \{\{r_i = 0.0\} \cup \{r_j = s_j^0 \text{ for } j \in 1..N, j\ne i\}\}
\end{align*}
Then the IP of node $i$ is computed as:
\begin{align*}
 IP_i= R(s_i^1) - R(s_i^0)
\end{align*}
The Improvement Potential measure serves as a clear indicator of which nodes might offer the greatest gain in minimizing risk. Acknowledging that total mitigation of all risk is unrealistic, a suitable pragmatic minimal risk greater than zero might be substituted in the above $s_i^1$ vector.
\item{\textbf{Birnbaum Importance:}} The BI measure considers the sensitivity of system reliability to the reliability of some component. The measure is therefore a partial derivative of system reliability, i.e., \vspace{-0.05in}
\begin{align*}
BI_i = \frac{\partial R(r)}{\partial r_i}
\end{align*}
Calculation of the BI measure can be conducted using the system risk function along with two state vectors, as follows.
\begin{align*}
s_i^0= \{\{r_i = 0.0\} \cup \{r_j | r_j \in [0,1]  \text{ for } j \in 1..N, j \ne i \}\} \\
s_i^1= \{\{r_i = 1.0\} \cup \{r_j | r_j \in [0,1]  \text{ for } j \in 1..N, j \ne i \}\} 
\end{align*}
Then BI measure is computed as:
\begin{align*}
 BI_i= R(s_i^1) - R(s_i^0) 
\end{align*}
\end{itemize}
With the measure of system risk and relative component importance values, it is possible to understand sources of significant risk in a system. Calculation of these values with differing supplier trust values will further yield insight into risks posed by suppliers.

\section{Case Study: Autonomous Vehicle}
In this section, we evaluate our proposed RIoTS methodology using a case study on a simplified model of an autonomous vehicle system.

\begin{figure}[H]
    \centering
    \includegraphics[width=.3\textwidth]{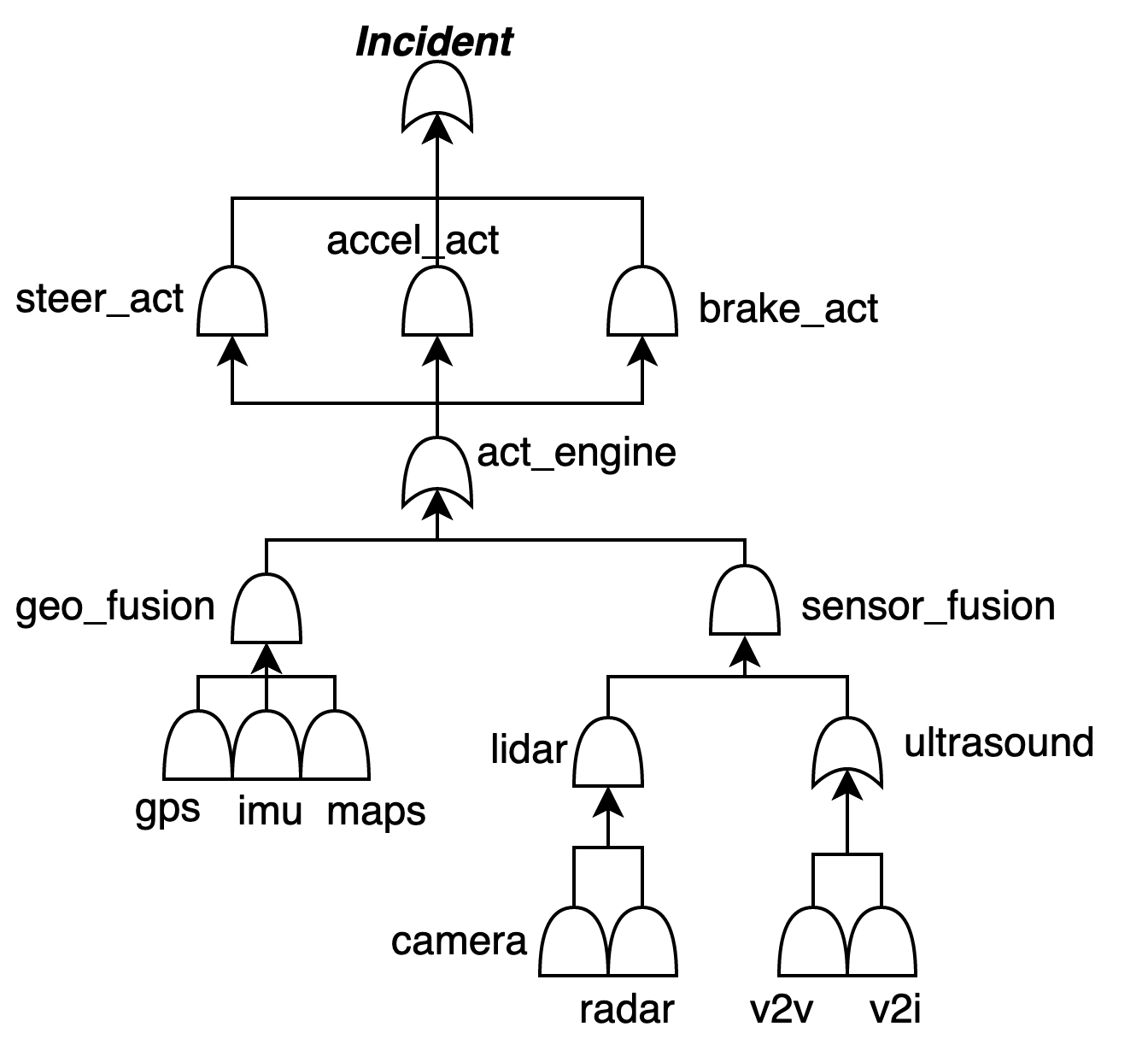}
    \caption{System graph for a simplified model of autonomous vehicle comprising of \emph{and} and \emph{or} nodes. \vspace{-0.05in}}
    \label{fig:at_sys}
\end{figure}

\begin{figure}[h]
    \centering
    \includegraphics[width=.4\textwidth]{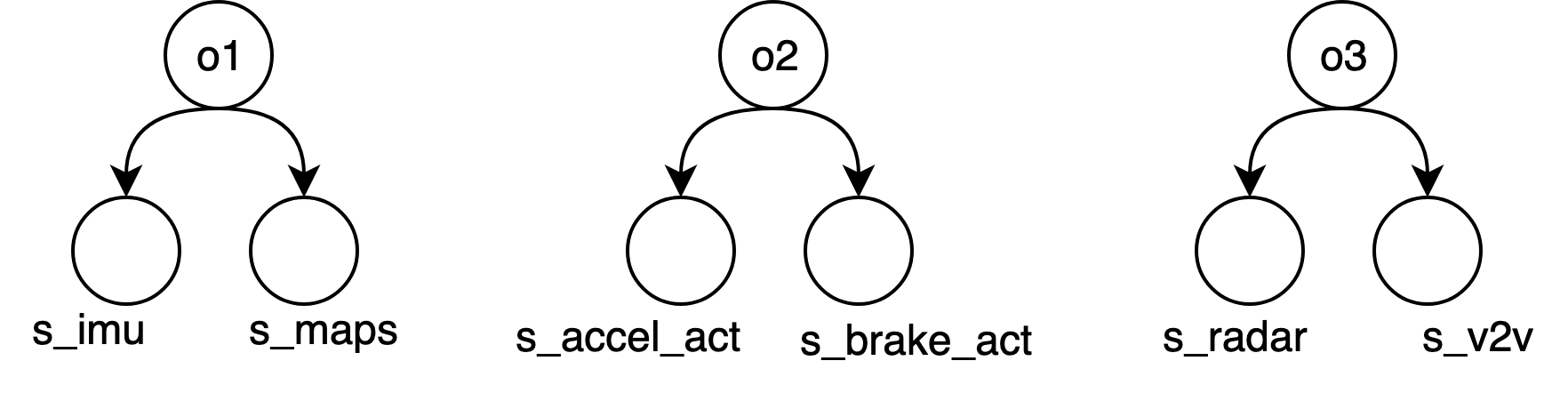}
    \caption{Supplier topology for autonomous vehicle system graph. Each owner controls a small group of suppliers.\vspace{-0.05in} 
    }
    \label{fig:s_top}
\end{figure}

 The related graph of components and their inter-dependencies is shown in Fig.~\ref{fig:at_sys}, with nodes depicted as and/or logic symbols. Each component has an associated supplier, denoted with the prefix ``s\_". For this simple case study, all suppliers are independent except the groupings shown in Fig.~\ref{fig:s_top}. These three groupings of suppliers will be useful to contrast the varying impact of supply chain risk according to structural importance of the related components.

\begin{figure*}%
\vspace{-0.2in}
    \centering
    \subfloat[Case 1 Risks]{{\includegraphics[width=2.3in]{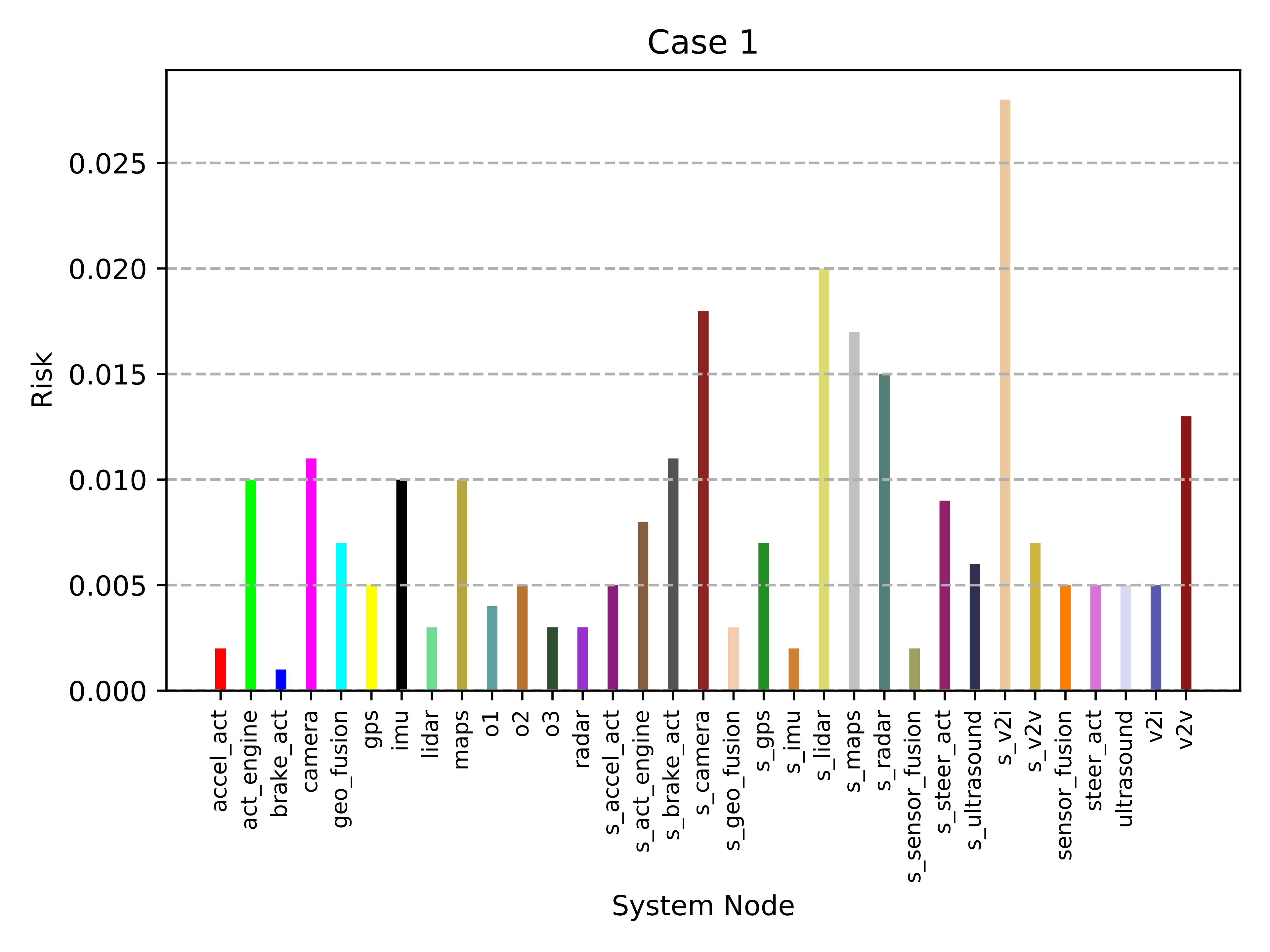}}\label{fig:case1_risk}}%
    \subfloat[Case 1 Improvement Potential]{{\includegraphics[width=2.3in]{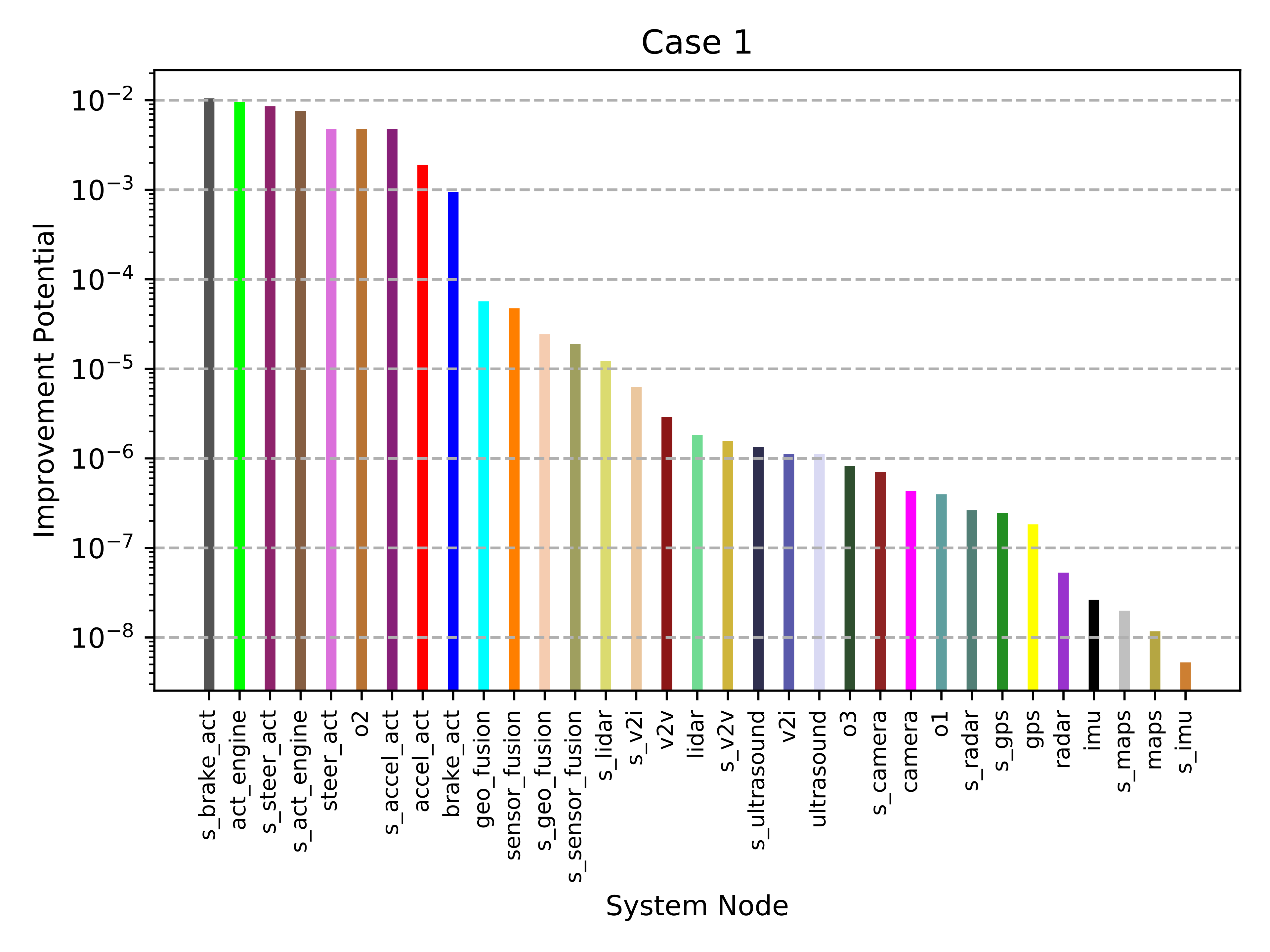} }\label{fig:case1_ip}}%
    \subfloat[Case 1 Birnbaum Importance]{{\includegraphics[width=2.3in]{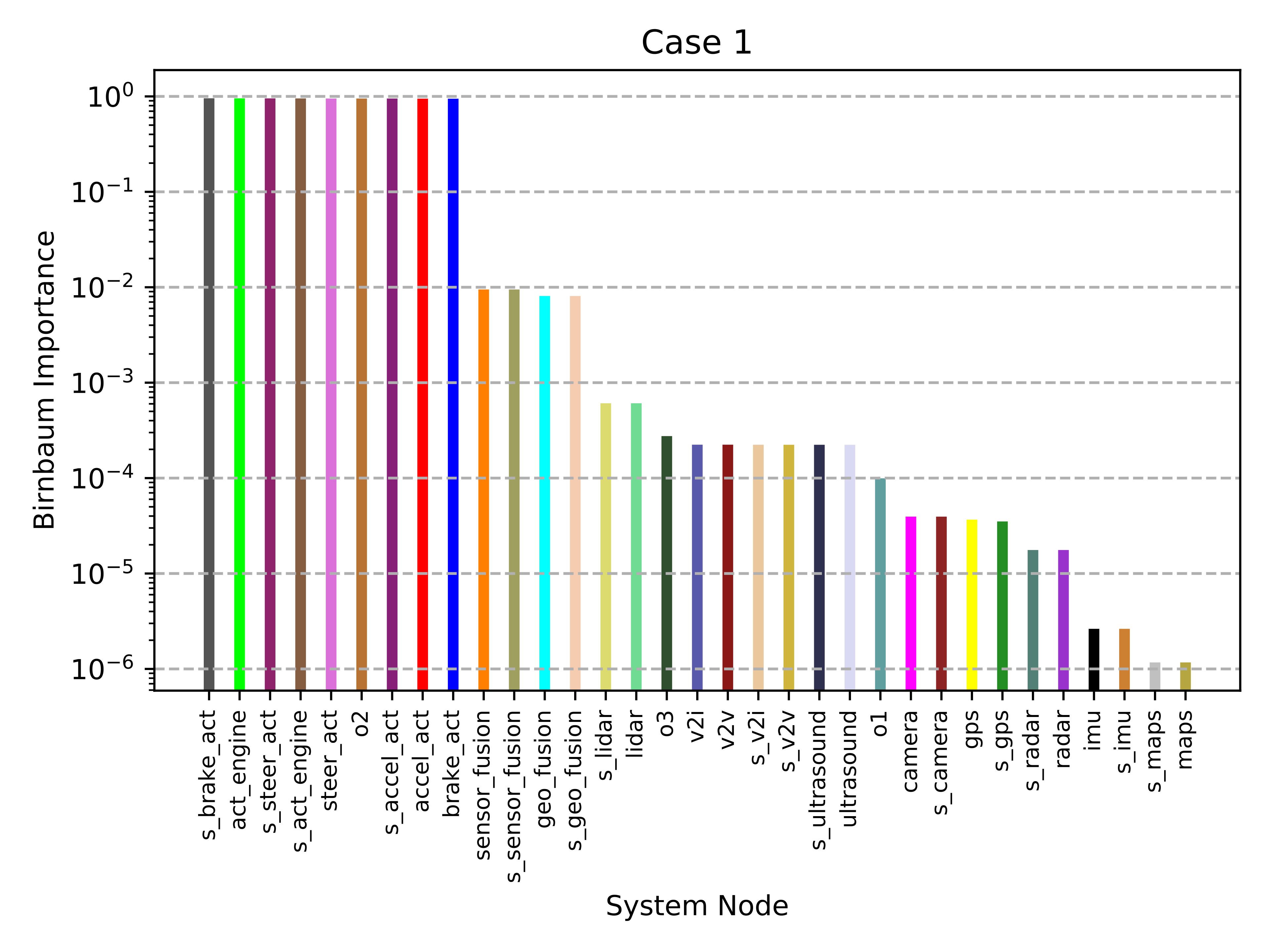} }\label{fig:case1_bi}}
    \caption{Autonomous vehicle risk analysis baseline case. Risk values and component importance measures are shown for suppliers, components, and supplier group controllers.}%
    \label{fig:case1}%
\end{figure*}

\begin{figure*}%
\vspace{-0.2in}
    \centering
    \subfloat[Case 2 Risks]{{\includegraphics[width=2.3in]{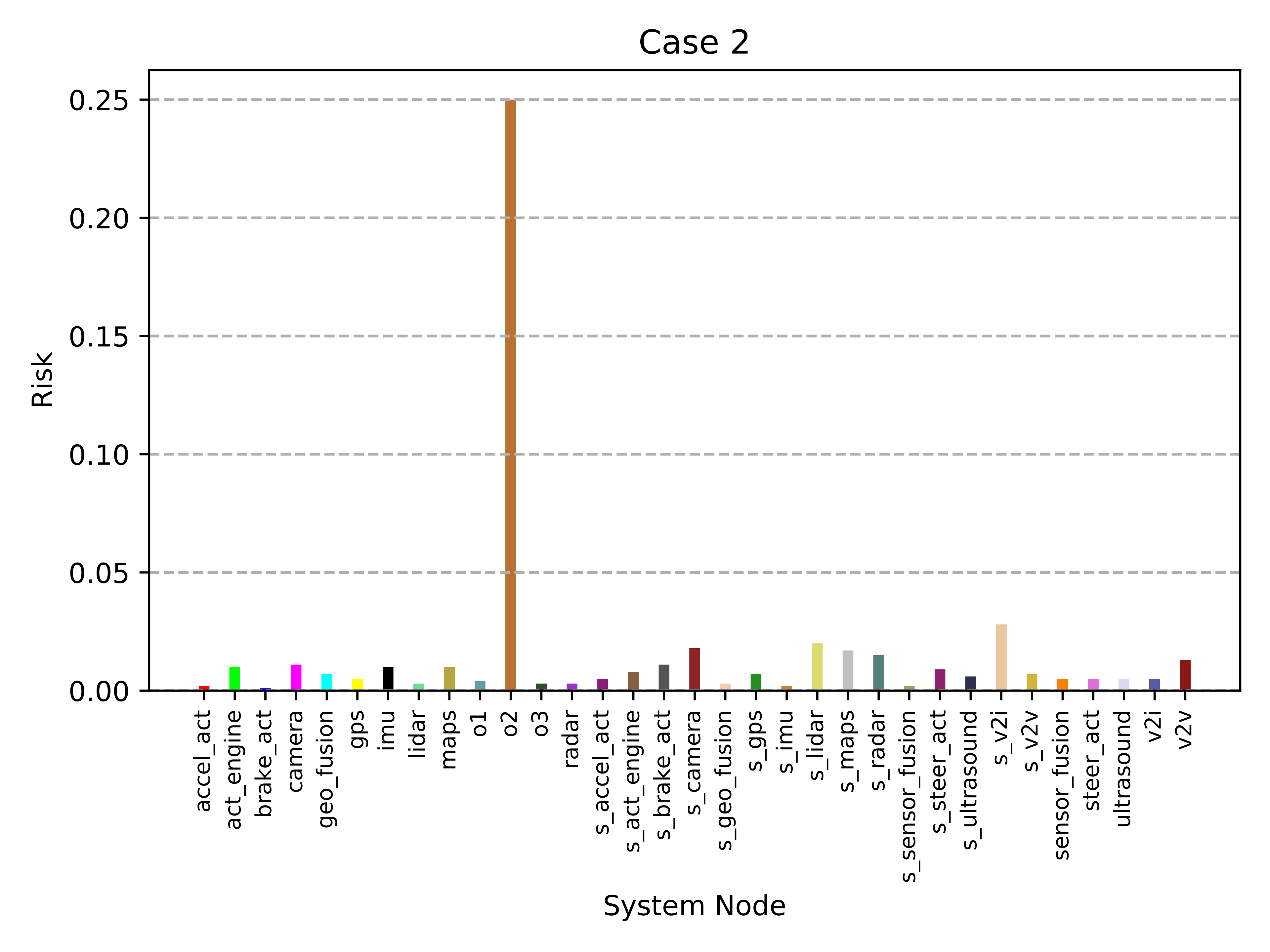}}\label{fig:case2_risk}}%
    \subfloat[Case 2 Improvement Potential]{{\includegraphics[width=2.3in]{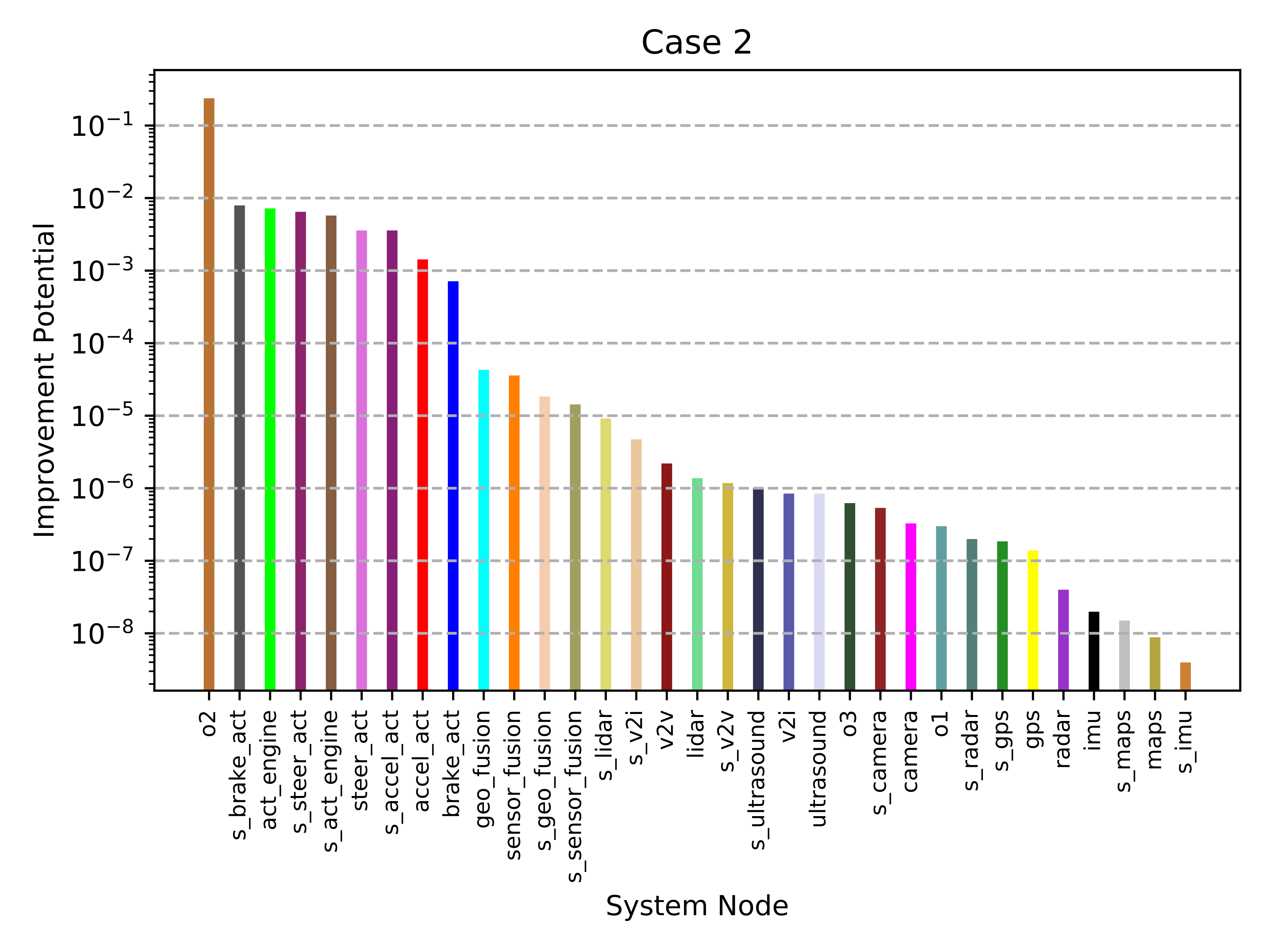} }\label{fig:case2_ip}}%
    \subfloat[Case 2 Birnbaum Importance]{{\includegraphics[width=2.3in]{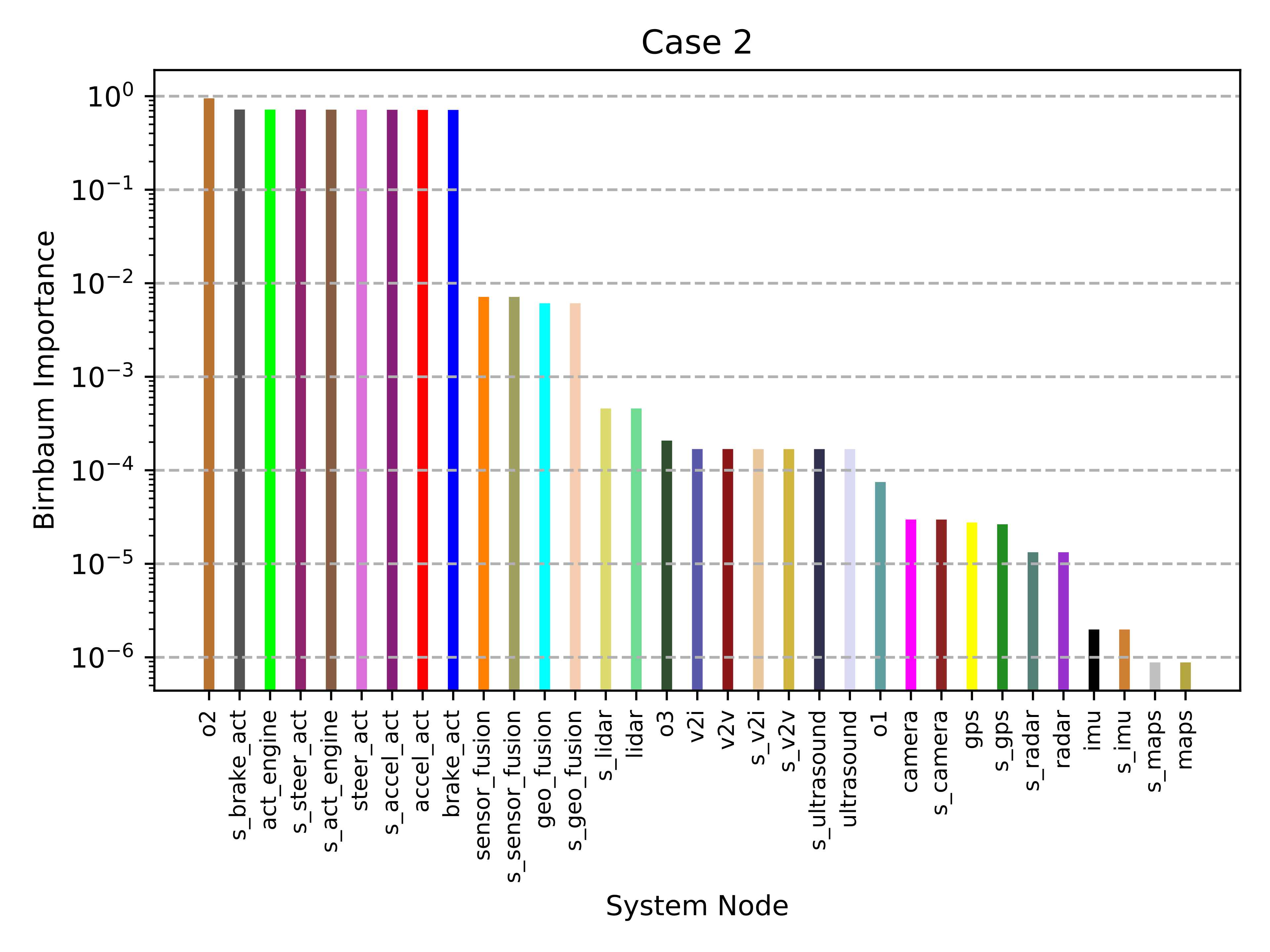} }\label{fig:case2_bi}}
    \caption{Autonomous vehicle risk analysis with one high risk supplier group posing significant system risk.}%
    \label{fig:case2}%
\end{figure*}

\begin{figure*}%
\vspace{-0.2in}
    \centering
    \subfloat[Case 3 Risks]{{\includegraphics[width=2.3in]{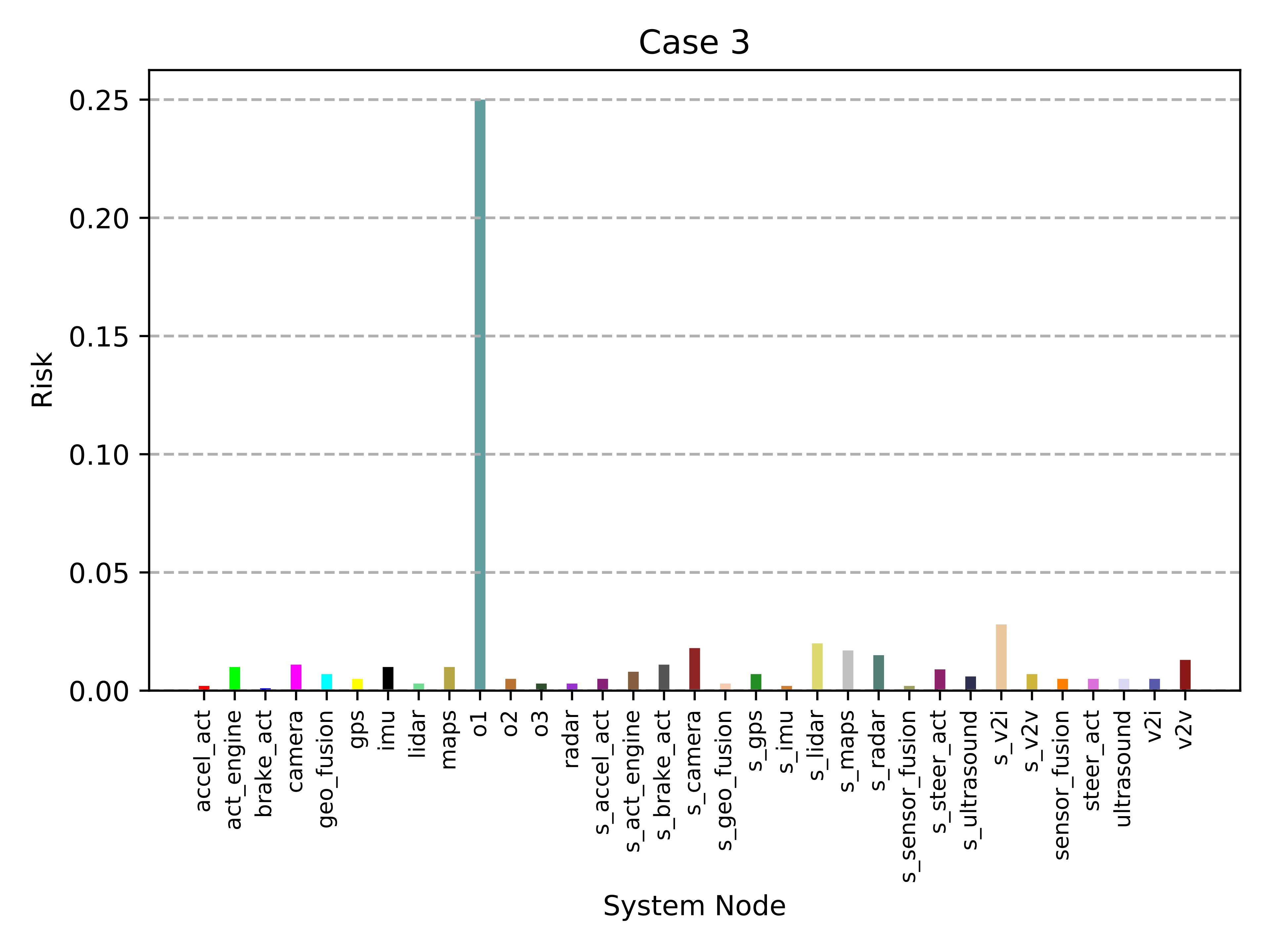}}\label{fig:case3_risk}}%
    \subfloat[Case 3 Improvement Potential]{{\includegraphics[width=2.3in]{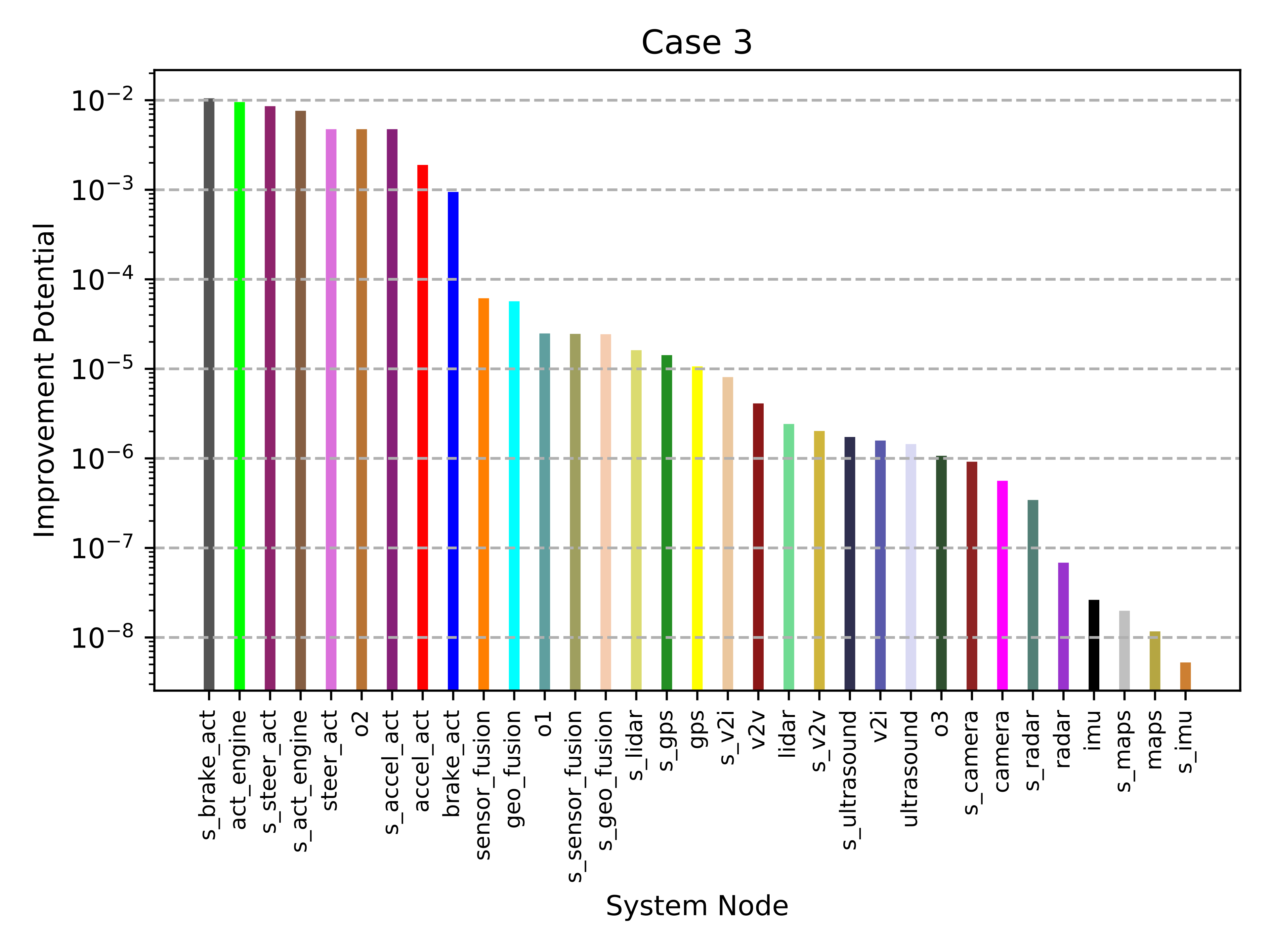} }\label{fig:case3_ip}}%
    \subfloat[Case 3 Birnbaum Importance]{\includegraphics[width=2.3in]{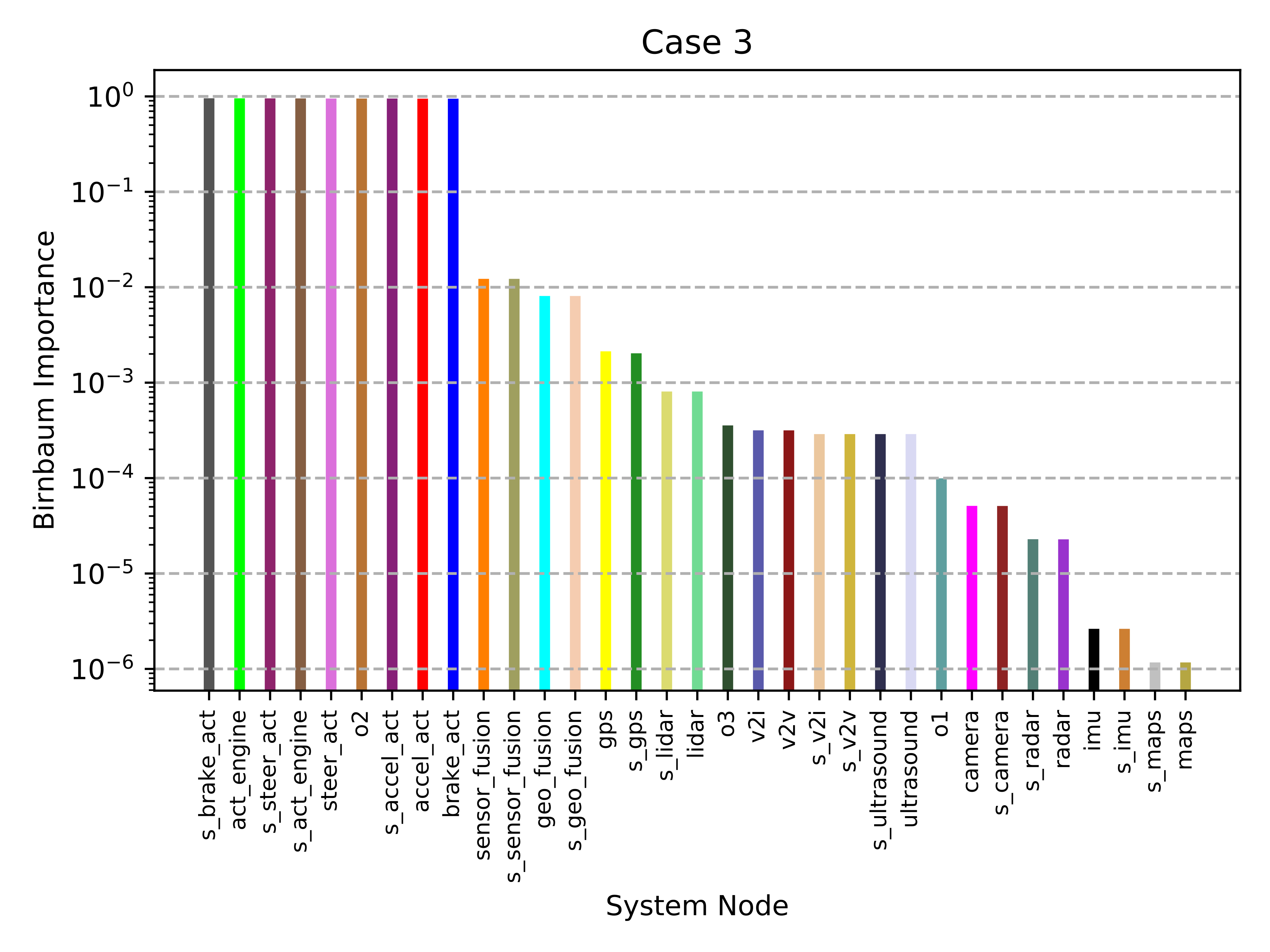} }\label{fig:case3_bi}
    \caption{Autonomous vehicle risk analysis with one high risk supplier group posing little system risk. \vspace{-0.1in}}%
    \label{fig:case3}%
\end{figure*}


This system will be subject to three different sets of risk values for the nodes. Case 1 provides a baseline scenario, with risks shown in Fig.~\ref{fig:case1_risk}. All risk values are below 0.03. When evaluating the systemic risk according to the methods described above, the result is 0.05475. This number was calculated by evaluating the probability of the various minimal cutsets that represent possible failure states, using~\eqref{equation:risk}.


Fig.~\ref{fig:case1_ip} and Fig.~\ref{fig:case1_bi} depict the Improvement Potentials and Birnbaum Importances of each node in the system. Several observations can be made about these measures. First, in this case, the two measures yield very similar orderings of system nodes. In our findings, this is not always the case, and depends on the system graph topology as well as the similarity between the risk values. Second, it is clear that the most important nodes by either measure are those nodes that are themselves minimal cutsets; in other words, those nodes that are single-points of failure for the system. This topological feature makes a much greater difference in importance than risk, as can be seen by comparing, for example, ``s\_v2i" and ``brake\_act". The former possesses high risk but low structural significance, and so ranks relatively low in importance measures. By contrast, despite having the lowest risk, ``brake\_act" ranks near the top in importance.


Case 2 begins with the risk values of case 1 but raises the risk of ``o2" to 0.25, as shown in Fig.~\ref{fig:case2_risk}. The system risk in this case was 0.28750, significantly higher than Case 1. Fig.~\ref{fig:case2_ip} and Fig.~\ref{fig:case2_bi} feature the respective importance measures. Here again we see similar orderings between the two measures. However, the Improvement Potential measure shows more significant divergence between the risky supplier and the other nodes.


Finally, Case 3 conducts a similar modification of ``o1" to be 0.25, returning ``o2" to its initial state. In this case, system risk is 0.05478, only slightly higher than Case 1, illustrating again the significance of the topological structure of the system in the calculation of risk. The importance measures for Case 3 show that the risky supplier ranks very low in importance. Once again, despite being a risky supplier with several dependent suppliers, the risk introduced to the system is very low compared to other more important nodes.

\vspace{-0.1in}
\section{Conclusion}
In this paper, we present RIoTS, a methodology to analyze risks in networked systems such as the IoT that emanate from the suppliers of individual components. Due to the comparatively unconstrained nature of the threat potentially posed by a malicious or compromised supplier, risk analysis must shift from a vulnerability-centered approach to the modeling of suppliers and components as a system. While attack tree techniques provide a foundation for our methodology, these techniques must be adapted to include suppliers and supplier groupings. The inclusion of supplier trust is grounded in the role played by suppliers in risk analysis procedures. Borrowing from established methods in reliability analysis, the use of minimal cutsets and importance measures provide measures of risk across a system and its individual components. While such system-level analysis involves computational complexity, future work will study the use of heuristics to achieve feasibly optimized risk mitigation.

\small
\vspace{-0.1in}
\bibliographystyle{IEEEtran}
\bibliography{references}

\end{document}